\documentclass{optica-article}

\journal{opticajournal} 

\articletype{Research Article}

\usepackage{lineno}
\usepackage{graphicx}%
\usepackage{multirow}%
\usepackage{mathrsfs}%
\usepackage[title]{appendix}%
\usepackage{xcolor}%
\usepackage{textcomp}%
\usepackage{manyfoot}%
\usepackage{booktabs}%
\usepackage{algorithm}%
\usepackage{algorithmicx}%
\usepackage{algpseudocode}%
\usepackage{listings}%
\usepackage{physics}
\usepackage[version=4]{mhchem}

\begin{document}

\title{The shadow of a laser beam}

\author{Raphael A. Abrahao,\authormark{1,2,\dag} Henri P. N. Morin,\authormark{1,3} Jordan T. R. Pagé,\authormark{1,4} Akbar Safari,\authormark{1,5} Robert W. Boyd, \authormark{1,4,6} and Jeff S. Lundeen\authormark{1,7,*}}

\address{\authormark{1}Department of Physics, University of Ottawa, 25 Templeton Street, Ottawa, Ontario, K1N 6N5, Canada\\
\authormark{2}Brookhaven National Laboratory, Upton, NY, 11973, USA\\
\authormark{3}Institute for Quantum Computing, Department of Physics and Astronomy, University of Waterloo, Waterloo, Ontario, N2L 3G1, Canada\\
\authormark{4}School of Electrical Engineering and Computer Science, University of Ottawa, Ottawa, ON, K1N 6N5, Canada\\
\authormark{5} Department of Physics, University of Wisconsin-Madison, Madison, WI, 53706, USA\\
\authormark{6}Institute of Optics, University of Rochester, Rochester, NY 14627, USA\\
\authormark{7}Joint Centre for Extreme Photonics, University of Ottawa - National Research Council of Canada, 100 Sussex Drive, Ottawa, Ontario K1A 0R6, Canada}

\email{\authormark{\dag}rakelabra@bnl.gov} 
\email{\authormark{*}jlundeen@uottawa.ca}

\begin{abstract*} 
Light, being massless, casts no shadow; under ordinary circumstances, photons pass right through each other unimpeded. Here, we demonstrate a laser beam acting like an object -- the beam casts a shadow upon a surface when the beam is illuminated by another light source. We observe a regular shadow in the sense it can be seen by the naked eye, it follows the contours of the surface it falls on, and it follows the position and shape of the object (the laser beam). Specifically, we use a nonlinear optical process involving four atomic levels of ruby. We are able to control the intensity of a transmitted laser beam by applying another perpendicular laser beam. We experimentally measure the dependence of the contrast of the shadow on the power of the laser beam, finding a maximum of approximately of approximately 22\%, similar to that of a shadow of a tree on a sunny day. We provide a theoretical model that predicts the contrast of the shadow. This work opens new possibilities for fabrication, imaging, and illumination.
\end{abstract*}

\section{Introduction}

Human's physical understanding of shadows developed hand-in-hand with
our understanding of light and optics. Throughout this multi-millennia
history, humans saw that shadows were cast by material objects like
trees, clouds, or the Moon. The study and use of shadows runs through the history of arts and science.
In theatre, shadow puppetry has existed for millennia in various cultures around the world. In fine art, the Renaissance
investigation of shadows, e.g., by Leonardo da Vinci and Dürer, contributed
to the development of perspective and realism in Western painting.
In astronomy, shadows revealed our relation to the cosmos; the discovery
that eclipses are shadows led to the measurement of the sizes and
distances of the Moon and the Sun, while Eratosthenes used shadows to
measure the circumference of the Earth. 

In optics, Ibn Al-Haytham's description of shadows lent credibility to the ray model of light, while
Grimaldi, Fresnel, and Arago's detailed observations of and predictions
for shadows introduced the concept of diffraction and the wave model.
In mathematics, the understanding that shadows are projections (i.e., silhouettes)
of an object led to early mapping-projections and seeded key concepts
in linear algebra. In philosophy, one of the most well-known texts is Plato's allegory of the cave, a discussion based on shadows. In technology, shadows are the basis of contact-lithography,
X-ray images, and a plethora of ways to measure and reconstruct three-dimensional
objects, e.g., computer-aided tomography. Over the century-spanning
course of this study, a shadow has come to be defined as a dark area
on a surface where light has been blocked by an object. All along,
regardless of whether the object is solid, gaseous, or liquid, or
whether it is opaque or translucent, it has been implicitly assumed
that it is material, i.e., made of some ordinary physical object with mass.

\begin{figure}[]
\includegraphics[width=1.0\textwidth]{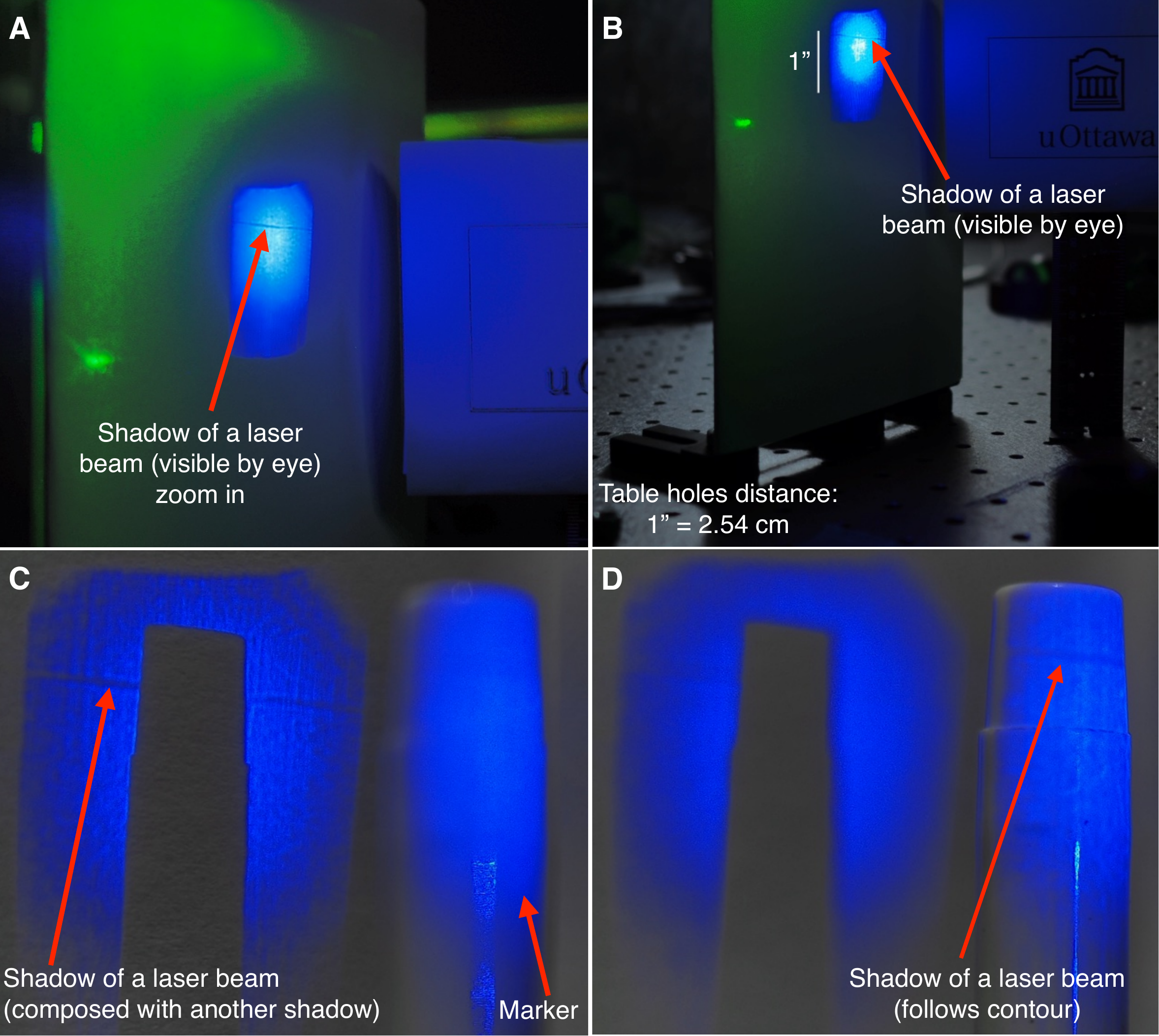} \caption{\textbf{Photographic images of the shadow of a laser beam}. A high-power green laser beam (the object), travelling through a cube of ruby, is illuminated from the side by blue light. (\textbf{A}) A photograph of the shadow cast by the object laser beam on a piece of white paper, image magnified approximately by a factor of 4 using a simple lens. The ruby cube length is about 1.2 cm, and the magnified image is about 4.8 cm. Hence, regardless of magnified or not, it portrays what can be seen in person. (\textbf{B}) A photographic image showing the surrounding for reference of scale. A white plastic marker (i.e., a broad tip pen) is placed in the path of the shadow, between the object beam and the paper, and the camera focus is fixed on (\textbf{C}) the paper or (\textbf{D}) the marker, thereby showing that the shadow follows the contours of the surface the shadow falls on. All images were taken with a regular consumer digital camera in a darkened room.}
\label{real_image} 
\end{figure}

We demonstrate a counter-example to that, a shadow cast by a laser beam. The shadow we observe has the usual characteristics of a typical shadow created
by a material object. Namely, the shadow fulfills the criteria: (i)
it is a large-scale effect that is (ii) visible by eye on ordinary
surfaces. Moreover, (iii) it is due to the object blocking the illumination
light and, thus, (iv) it takes the shape of the illuminated object
and (v) follows as the object changes position or shape. Lastly, (vi)
the shadow follows the contour of the object it falls on, giving the
sense of three-dimensionality (3D) that interested da Vinci and creating
an effect that is often used for depth measurement in the 3D imaging
of scenes by cameras.

As our object we use a laser beam, namely a high-power
green beam with an optical wavelength of 532 nm. This object beam
travels through a cube of standard ruby crystal. We illuminate the
beam from the side with blue light. We start with qualitative observations.
Fig.~\ref{real_image}~(A) and (B) show the shadow of the object beam cast
on a piece of paper. The laser shadow extends through an entire face of the rube cube, about 1.2 cm long, confirming the macroscopic scale, criteria (i). Moreover, Fig.~\ref{real_image}~(A)
is a photo taken with a consumer photography camera. It genuinely
portrays what can be seen in person, criteria (ii). Fig.~\ref{real_image}~(C)
and (D) show the shadow falling across a marker made from white plastic
that is placed in front of the paper. Just as one would expect, the
shadow follows the depth-contours of the scene: the rounded surface
of the marker and the flat paper behind it, thereby confirming criteria
(vi). Thus, three of the six criteria have been qualitatively confirmed. 

Under ordinary circumstances, photons do not interact with each other, much less block each other as needed for a shadow. However, photon-photon interactions can happen in limited situations. At extreme optical intensities, it has been predicted that the field can polarize the vacuum and thereby affect another field~\cite{scattering_vacuum_1,scattering_vacuum_2,scattering_vacuum_3,scattering_vacuum_4,scattering_vacuum_5}. In specially prepared gases (e.g., of Rydberg atoms), large atomic dipoles have been demonstrated to mediate photon interaction~\cite{Rydberg1,Rydberg2,Rydberg3}. A combination of bosonic statistics and quantum interference can lead to photons bunching together, e.g., in Hong-Ou-Mandel interference~\cite{HOM_effect,bouchard_HOM_review_2020,PRA_Jordan2022}. The laser shadow effect arises from the diverse subject area of nonlinear optics~\cite{boyd_nonlinear}, in which matter reacts nonlinearly to an applied optical field and can then influence another field. 

In the next section, we propose a physical model for the mechanism in the ruby that underlies the laser shadow effect. We present a comparison of the quantitative predictions of the model with measurements of the shadow’s shape and contrast. We then return to the remaining two criteria and provide evidence that they too are satisfied by the laser shadow effect. Lastly, before concluding, we discuss alternative interpretations of the effect and highlight some key differences between the laser shadow effect and other nonlinear optical effects.

\begin{figure}[]
\centering \includegraphics[width=1.0\textwidth]{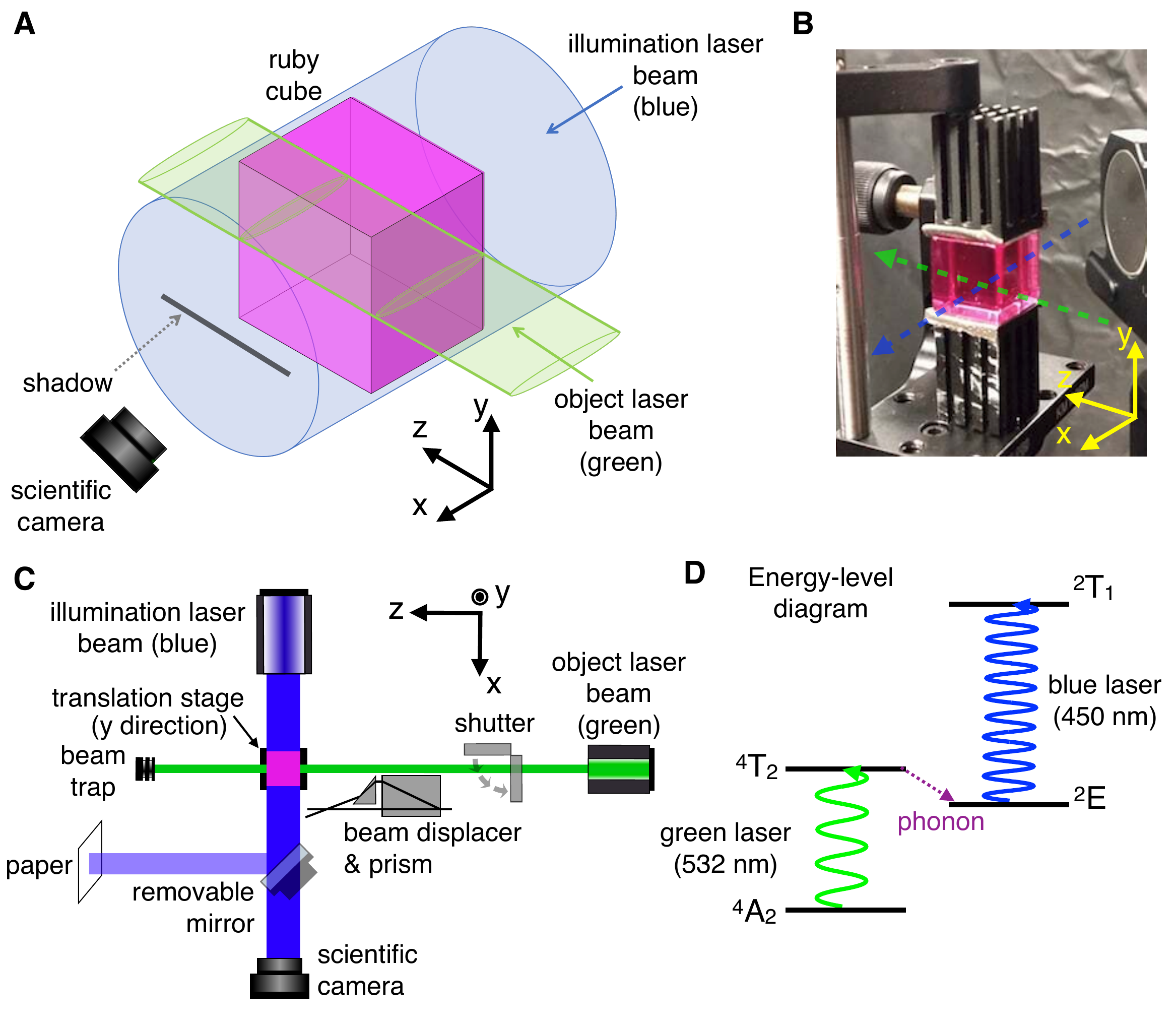} \caption{\textbf{Scheme to create and observe the laser shadow.} \textbf{(A)}
The object laser beam (green) travels through a ruby cube and casts
a shadow in the illuminating blue light. \textbf{(B)} A photograph of the
ruby cube overlayed with the beam directions. (\textbf{C})
The simplified experimental setup to observe the shadow cast either on paper
or, for quantitative measurements, impinging directly on the camera.
(\textbf{D}) The relevant energy level diagram of ruby.
As described in detail in the Theory Section, a photon from the green
laser beam excites the lower transition, which then causes electrons to populate the $^{2}\text{E}$ energy level, and consequently increases the absorption (i.e., blocks) of the blue (illumination) light.}
\label{concept} 
\end{figure}

\section{Theory}

In order to observe the shadow of a laser beam, one requires a medium that exhibits a strong nonlinear absorption. However, most materials exhibit saturation of absorption at high laser intensities, meaning that the material becomes more transparent in the presence of a strong laser field. This would lead to an ``anti-shadow" where the shadow location of the laser beam appears brighter than the background. 

However, some materials can exhibit an increase in absorption at higher intensities under certain conditions. This response is known as reverse saturation of absorption and requires particular conditions which includes a more than two-level system. Moreover, the excited state must have a larger absorption cross-section than the ground state. In addition, neither the first
nor the second excited states should decay to other levels
that can trap the atomic population. Furthermore, the incident
light should saturate the first transition only. A recent study shows that ruby satisfies all these conditions and exhibits reverse-saturation of absorption~\cite{Safari_arxiv2023}.

The laser shadow effect is conceptually depicted in Fig.~\ref{concept}~(A)--(D),
where we show the propagating paths of the object laser beam (green)
and the illumination laser beam (blue). The laser shadow effect is a consequence
of the optical nonlinear absorption (i.e., reverse-saturation of absorption, equivalently called saturable transmission)
in the ruby; wherever the object laser beam (green) exists in the
ruby, it increases the optical absorption of the illuminating laser beam (blue). This results in a matching region in the illuminating light with lower optical intensity, a darker area that is the shadow of the green laser beam.

At a fundamental level, this effect is explained by the atomic structure
of the ruby and its optical properties \cite{boyd_nonlinear,boyd_gaeta2008,bigelow2003superluminal,gehring2006observation,kramer1986nonlinear,bigelow2003observation,franke2011rotary,cronemeyer1966optical,Maiman_PRL_1960,maiman1960stimulatedNature,boyd_gauthier2002,boyd2009controlling,Kiang1965}.
Ruby (\ce{Al2O3}:\ce{Cr}) is an aluminium oxide crystal and the distinct
ruby-red colour comes from chromium impurities (i.e., atoms) that
distort the crystal lattice. The relevant energy levels of the ruby crystal lattice presented in
the diagram in Fig.~\ref{concept}~(D) exhibit an unusual interaction,
that we will now describe, between select colours of light. One colour
is green, the object laser beam, which has an optical wavelength of
532 nm. It will drive the transition from the ground state $^{4}\text{A}_{2}$
to an excited state $^{4}\text{T}_{2}$, which then decays rapidly via phonons
to the $^{2}\text{E}$ state. This then allows the electrons
to absorb blue light (450 nm), the illumination, by transitioning
from $^{2}\text{E}$ to $^{2}\text{T}_{1}$. However, the
blue laser (450 nm) could in principle be absorbed by the electrons in $^{4}\text{A}_{2}$ and transition to some other level (not shown in the diagram). The effect will
only take place if the absorption cross-section of the second transition ($^{2}\text{E}$
to $^{2}\text{T}_{1}$) is larger than the one of the first transition ($^{4}\text{A}_{2}$
to $^{4}\text{T}_{2}$), which is the special case exhibited by ruby.

Two comments about the observation of the laser shadow effect need to be made explicitly. First, both lasers, the green and blue, are not on resonance to each other, as this is not required to achieved the effect. Second, the blue laser transition does not cycle. These two aspects make it simpler to use and explore the laser shadow effect.

In the Theoretical Simulations in the Supplement 1, we present the analytical rate equations
that model these transitions and, thus, model the laser shadow effect.
This model shows that the contrast between the shadow and its surrounding
illumination increases monotonically with optical power of the green
laser and that the shape of the shadow follows the spatial intensity
profile of the green laser beam. In the next section, we present our quantitative analysis of the laser shadow effect and the comparison between measurements and the model.

\section{Experiment and Results}

\label{sec:exp_res}

\begin{figure*}
\centering \includegraphics[width=0.5\textwidth]{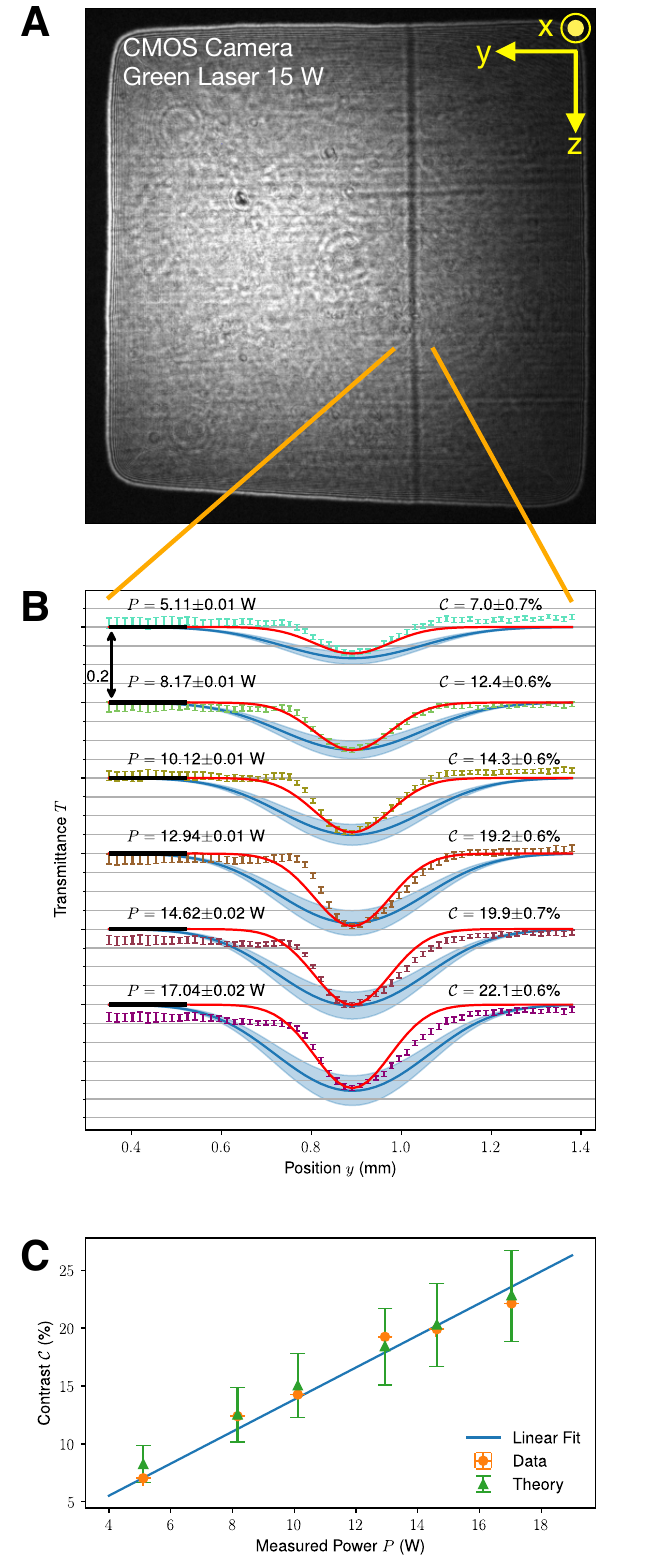} \caption{\textbf{Quantitative analysis of the shadow contrast.} (\textbf{A})
Direct image of the illuminating light transmitted through the ruby
for a 15 W object laser beam. The vertical line of lower brightness
is the shadow of the green laser beam. (\textbf{B})
For the 1 mm wide y-region in (A) indicated by the orange lines, the
experimental (dots, error bars are the standard error from 21 trials)
and theoretical (solid blue line and shaded error bands from the model
in the Theoretical Simulations) relative transmittance $T$ through
the ruby cube of the blue illuminating light is plotted for six object-laser
optical-powers. For clarity, the six datasets are separated vertically
by $0.2$ alongside their respective power $P$ of the green object
laser and contrast $\mathcal{C}$ (Eq.~\ref{eq_contrast}). For each,
the horizontal black solid line at the left marks a transmittance
of $T=1$ relative to the transmittance when the object laser is absent.
The solid red line is the Gaussian fit of the measured laser beam
spatial profile, which shows that the shadow shape is the same as
the object laser spatial profile. (\textbf{C})
Peak contrast (experiment: orange circles with plotted but not visible
error bars in both $\mathcal{C}$ and $P$; theory: green triangles
with error bars) for the six power values in (B) along with a linear fit with zero-intercept.}
\label{analysis_graph} 
\end{figure*}

\begin{figure*}
\centering \includegraphics[width=1\textwidth]{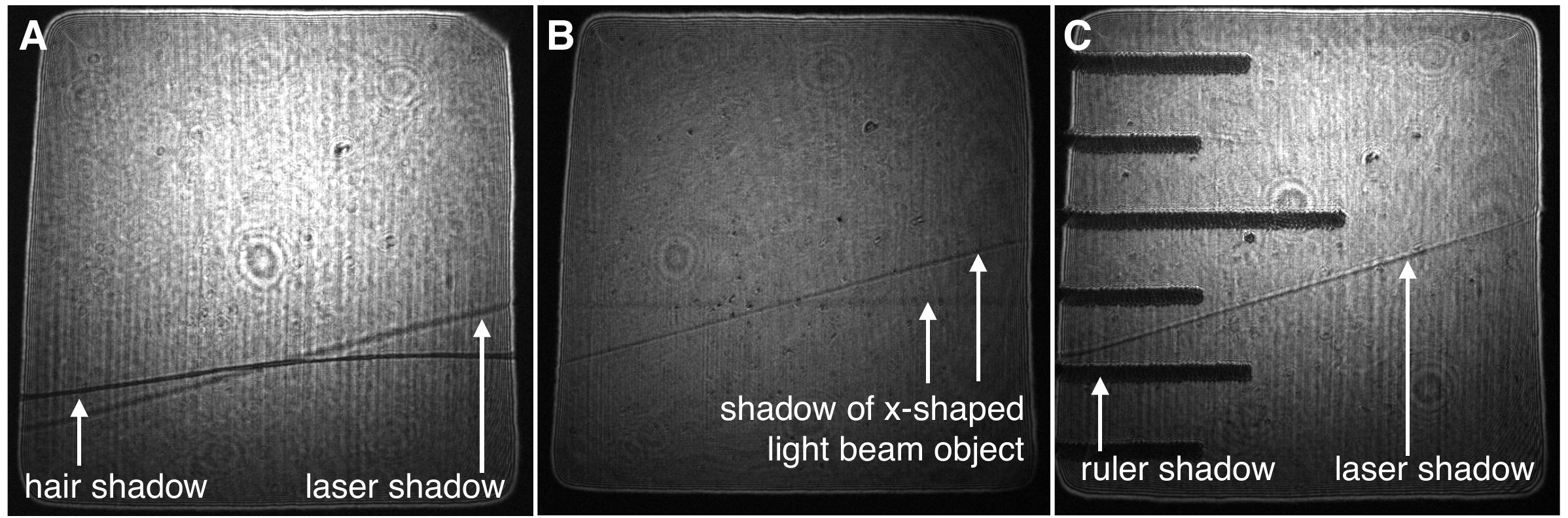} \caption{\textbf{Qualitative comparison of the laser shadow to normal shadows.}
Three direct images of the transmitted illuminating light containing
multiple simultaneous shadows. (\textbf{A}) A human
hair produces a very similar shadow to that of the object laser beam
($P=20$ W). (\textbf{B}) The shadow of an object
made of two crossed laser beams, showing that the laser shadow has
the same shape as the object (total $P=20$ W). (\textbf{C}) For
scale, a shadow of the object laser beam ($P=15$ W) with a ruler
(imperial, small increments are $1/6"=1.588$ mm).}
\label{figextra_1} 
\end{figure*}

A simplified schematic of the experimental setup is presented in Fig~\ref{concept}~(C)
and a detailed schematic is given in the Supplement 1 along
with a comprehensive description. We outline the key elements of the
experiment here. A continuous wave (CW) laser diode creates the blue illuminating
light which is collimated and enlarged with lenses to fill the the
ruby cube, which has an edge length of $12.0\pm0.5$ mm and a doping of $2.5 \cross 10^{25}$ m$^{-3} \pm 10\%$. The green object laser is produced by a CW diode-pumped solid-state
laser. The theoretical model shows that the local opacity of the green
laser beam saturates at high intensities, i.e., any chosen point in the
laser beam object will be translucent to the blue illumination. By
increasing the object beam thickness along $x$, the direction along the illumination, we maximize the interaction length between the object and the illumination lights. To accomplish this, the object beam has
an elliptical profile, i.e., approximately Gaussian, with $1/e^{2}$
intensity half-widths of $w_{0x}=3.17\pm0.01$ mm and $w_{0y}=0.168\pm0.003$
mm in the $x$ and $y$ directions, respectively. Absorption of the
green laser heats the cube, which changes the phonon population and
lattice spacing and, in turn, the opacity. Consequently, we limit
the heating time as needed to keep the cube's temperature low, which
we monitor. Images were taken in two different ways. The first was
designed to capture what is seen by eye. In it, a mirror reflects
the transmitted blue illumination towards a piece of paper, which is then photographed using a digital consumer camera and a lens (see Fig.~\ref{real_image},
discussed above, and the Supplement 1). In the second way, a scientific monochrome
camera is directly in the path of the transmitted blue light in order
to take quantitative data, which we present next.

We now describe how we quantitatively analyze these images and compare
them to our theoretical model. Figure \ref{analysis_graph}~(A) shows
a typical image obtained by the scientific camera when the object
laser power is set to $P=15$ W, clearly showing the laser shadow. We take a set of 21 images used for normalization without the object (green) laser, followed by 21 images
with the object (green) laser, and thus, shadow present. For all the images
we integrate along $z$, resulting in a distribution in $y$ only.
We divide each shadow-image distribution $s(y)$ by the normalization-image
distribution $n(y)$ to find the relative transmittance at each $y$
position, $T(y)\equiv s(y)/n(y)$. The mean relative transmittance
and its standard error over 21 images is shown in Fig.~\ref{analysis_graph}~(B).
In the absence of a shadow, nominally $T(y)=1$ at all $y$, whereas
a perfectly black shadow would have $T=0$. Continuing this procedure,
we take 21 images for each of six object laser powers spanning from $P=$ 5
W to 17 W (measured just before the cube). Fig.~\ref{analysis_graph}~(B)
shows the resulting experimental relative transmittance along with
the corresponding theoretical curves with error bands found from the
uncertainties in the parameters in the model (see Theoretical Simulations). Note that in Fig.~\ref{analysis_graph}~(B), it is possible that some experimental values of the normalized Transmittance $T$ could display values above 1.0. This is purely an artifact of the mathematical normalization based on a real image, no physical meaning is associated to that. To facilitate comparing between different configurations of object laser (green) optical power and for sake of clarity of the figure, each set of graphs for a given power is vertically shifted by 0.2 with a solid black line to indicate a reference level for the drop in transmittance according to that level.

The experimental relative transmittance reveals a few important characteristics
of the laser shadow effect. The experiment and theoretical model agree
well for the largest drop in transmittance, suggesting that our physical
model correctly predicts the maximum amount of illumination that is
blocked. However, the width of the theoretical shadow is approximately
twice that of the measured shadow. An unexpectedly narrow shadow could
potentially point to other non-shadow-like nonlinear optical effects
playing a role in the formation of the observed dark region. For example,
nonlinear optical refraction by the object beam could divert the illumination
rather than block it, creating dark regions much like those formed
at the bottom of a swimming pool by surface waves. To distinguish
between these possible causes, we compare the shadow shape to the
profile along $y$ of the object laser beam. A Gaussian fit (Fig.~\ref{analysis_graph}~(B),
solid red line) to the object beam's profile closely follows the experimental
profile, albeit not within error. This confirms the expectation that
the shadow should be the same as size as the object (criteria (iv))
and that it is caused by blocking rather than diverting the illuminating
light (criteria (iii)). Another possible hypothesis for the width discrepancy between measurements and the theoretical model is a self-focusing process as the object laser beam (green) goes through the ruby cube~\cite{hogan2023}.

The darkness of a shadow is its main characteristic and so we now
investigate this more thoroughly. More specifically, we compare the
darkness to the illumination using the contrast $\mathcal{C}$: 

\begin{equation}
\mathcal{C}\equiv\frac{T^{top}(y_{\mathrm{min}})-T^{bottom}\left(y_{\mathrm{min}}\right)}{T^{top}(y_{\mathrm{min}})},\label{eq_contrast}
\end{equation}
where, over all $y$, \textit{$T^{bottom}(y_{\mathrm{min}})$} is the minimum value of transmittance $T$ for the illumination laser beam (blue), i.e., darkest transmitted value, with the object laser beam (green) present, while $T^{top}(y_{\mathrm{min}})$ is the value of transmittance $T$ for the illumination laser beam (blue) at the same position, $y_{\mathrm{min}}$, when the object laser beam (green) is absent. Essentially, the contrast $\mathcal{C}$ is a normalized metric of the drop in transmittance of the illumination laser beam (blue).

The graph of the experimental contrast as a function
of optical power of the green laser is reported in Fig.~\ref{analysis_graph}~(C),
together with a linear fit ($\mathcal{C}=mP, m=1.38 \pm 0.03~\%/\mathrm{W}$)
and the theoretical predictions for each of the six powers. To evaluate the quality of the linear fit, we obtained $R^2 = 0.9669$ and $\chi^2 = 0.307$, which means that the linear distribution is a good fit of the observed data \cite{hogg1977probability}. The linearity
shows that the object beam is far from saturating the $^{4}\text{A}_{2}$
to $^{4}\text{T}_{2}$ transition, as was the goal of creating the
elliptical object beam. The maximum contrast achieved was $22.1\pm0.6$\%
with $17.04\pm0.02$ W in the object beam. For this case, we infer an
attenuation coefficient that the illuminating light experiences while
it travels through the object beam, $\alpha=173.6~\mathrm{m^{-1}}$ (the $\alpha$ in Eq.~(S7) in Supplement 1). We conclude that as the object laser beam power increases, the deeper the drop
in transmittance of the illuminating blue light, leading to a greater
magnitude of the laser shadow effect, which in turn is manifested as a
greater contrast value.

In this last part, we return to our qualitative observations, namely comparisons
of the laser shadow to a regular shadow made from a material object.
Fig. \ref{figextra_1} presents three direct images, taken with the
scientific camera, each showing a different comparison. In Fig. \ref{figextra_1}
(A), while the object beam is present, we place a strand of black
hair in the path of illuminating blue light before the cube. Unsurprisingly,
the black hair is more opaque than the object laser beam, which transmits
almost 80\% of the illumination, and casts a darker shadow. Nonetheless,
without the labels, it would be difficult to decide which shadow is
which. In Fig. \ref{figextra_1} (B), we split and divert the green
laser using a beam displacer and prism with the goal of changing the
light object to be x-shaped, i.e., two beams crossing at an acute
angle. As with the shadow of a material object, the laser shadow in
the image takes on the new shape of the object (e.g., as with shadow
puppets) confirming criteria (iv), i.e., the laser shadow is x-shaped.
In Fig. \ref{figextra_1} (C), a ruler, placed before the ruby cube, and the object laser are both present,
showing the length-scale of the effect (criteria (i)), and thus its macroscopic manifestation. To confirm the final criteria,
in the Supplement 1, we include a real-time video of one
of the object laser beam changing in angle and position. Just as criteria
(v) demanded, the shadow follows the object beam as it moves without
an observable delay by the human eye. The theoretical model presented in the Supplement 1 indicates that the response time of the laser shadow effect is in the millisecond regime.

The laser shadow effect requires a ruby to mediate this blockage, which raises the interesting question of whether the photons in the object laser themselves are blocking the illuminating light or rather it is the atoms in the ruby. A analogous question --- is a light wave propagating in a material made up of photons or excitations in atoms? --- applies to established effects such as slowed or stopped light and, indeed, even everyday transmission through a glass window. Fundamentally, the wave is actually composed of a hybrid of the two, polaritons. We are in the presence of strong absorption (hence the shadow) and the polaritonic nature of the excitation in the medium, and both concepts are necessary to understand what is happening. Strictly speaking, it is not massless light that is creating the shadow, but it is the material counterpart of the polariton, which has mass, that is casting the shadow. 

We now clarify how this shadow effect is distinct from other known nonlinear optical effects. An informative, if extreme, example is an all-optical switch in which the presence of a beam switches a second beam from output port A to B. The absence of the second beam at A is only a shadow in a very limited sense. Specifically, the switch will only operate for these two beams; for other beam angles, frequencies, bandwidths, durations, arrival times, and positions, all-optical switches cease to function well or, more likely, do not function at all. The switch is an engineered device that is optimized for narrowly specified inputs, far from what most people would call an object casting a shadow.

Other common nonlinear optical effects are similarly limited in scope. Electromagnetically induced absorption occurs for an extremely narrow bandwidth. Optical pumping causes an atomic population to be transferred into a new energy level that is resonant with a probe and thus increases its absorption. Hence, this is typically a narrowband effect requiring lasers that are frequency-locked to the atomic transition energies. For broadband light, the change in absorption is tiny. Optical phase-change materials can change their transmissivity upon optical excitation, but this is due to a change in the material structure itself and is, thus, slow for macroscopic systems typically. In summary, while there exist nonlinear effects in which light controls the transmission of light, they require precise and often elaborate engineering and only appear in specific geometries and wavelengths.
The light-light interaction in ruby that we demonstrate has unique characteristics that distinguish it from the above nonlinear effects.  Namely, the reverse saturation of absorption of ruby occurs for a broad range of wavelengths, they are in the visible spectrum, and the illumination can travel in any direction. Consequently, no specialized instrumentation or elaborate design was needed to demonstrate the absorption, which is strong enough to be easily visible by the naked eye. In our case, the effect is very broadband; the shadow can be seen even with a blue LED playing the role of illumination, rather than a laser. Prior to the present work, to the best of our knowledge, there is no other mechanism that can be used to show a laser shadow that is obvious and clear, broadband, and visible to the naked eye. None of the aforementioned cases could lead to something similar to the laser shadow effect as described in our work. In summary, the unusual properties of ruby make possible the first demonstration of light-light absorption that behaves like a regular shadow.

\section{Conclusion}

We showed how a laser beam can be made to cast a shadow that behaves
as any other ordinary shadow. To that point, the laser shadow obeys
six straightforward criteria that distinguish ordinary shadows from
other phenomena that are superficially similar, such as nonlinear
optical refraction, light-darkening glass, temperature-sensitive mirrors,
or laser-induced damage in glass. Of particular importance is that we found a successful agreement of our analytical physical model with our quantitative observations, showing that the mechanism behind the laser shadow is a blockage of the illuminating light. 

Other materials such as alexandrite are expected to also be able to show the laser shadow effect~\cite{malcuit1984saturation_alexandrite}.

This experiment redefines our understanding of what a shadow is --
under the right conditions, laser beams can cast a shadow. Potential applications can be envisioned in areas such as optical switching \cite{dawes2005all}, controllable shade or transmission, control of the opaqueness of light with light, and lithography. 

\begin{backmatter}
\bmsection{Funding}
This work was supported by the Canada Research Chairs (CRC) Program, the Natural Sciences and Engineering Research Council (NSERC), the Canada Excellence Research Chairs (CERC) Program, the Canada First Research Excellence Fund award on Transformative Quantum Technologies, the U.S. Department of Energy QuantISED award, and the Brookhaven National Laboratory LDRD grant 22-22. HPNM acknowledges support from the Canada Graduate Scholarships - Master's (CGS-M) program.


\bmsection{Disclosures}
The authors declare no conflicts of interest.

\bmsection{Data availability} Data underlying the results presented in this paper may be obtained from the authors upon reasonable request.

\bmsection{Supplemental document}
See Supplement 1 for supporting content. 

\end{backmatter}


\bibliography{references}

\end{document}